\documentclass{proceedings}
\usepackage[english]{babel}
\usepackage[latin1]{inputenc}
\usepackage{epsfig}
\usepackage{epsf}
\usepackage{multirow}
\usepackage{amssymb}
\usepackage{dcolumn}

\newcommand{\beq} {\begin{equation}}
\newcommand{\eeq} {\end{equation}}

\newcommand{\bqu}{{\bf q}}

\newcommand{\half}{\frac{1}{2}}
\begin{document}
\pagestyle{prochead}

\title{Short-range nucleon-nucleon correlation effects in \\
       photon- and electron-nucleus reactions}

\author{A.M. Lallena} \author{M. Anguiano} \affiliation
{Departamento de F\'{\i}sica Moderna, Universidad de Granada, 
 E-18071 Granada, Spain.}  
\author{G. Co'~}
\affiliation {Dipartimento di Fisica, Unversit\`a di Lecce and I.N.F.N.
Sez. di Lecce, I-73100 Lecce, Italy\\~\\}

\begin{abstract}
The role played by nucleon-nucleon short-range correlations in photon-
and electron-nucleus reactions is analyzed with a model which includes
all the diagrams containing a single correlation function. The
excitation of low-lying high-spin states, the reactions (e,e'),
(e,e'N) and ($\gamma$,N) and the two-proton emission by both electrons
and real photons are investigated. The results obtained for
doubly-closed shell nuclei show that ($\gamma$,pp) appears to be the
most adequate process to identify short-range correlations in the
clearer way.
\end{abstract}

\maketitle

\section{The model}

In these last years we have developed a model to describe
electromagnetic responses of nuclei with $A>4$ by considering also
SRC. Our model has been applied to describe inclusive (e,e') processes 
\cite{co98,mok00,co01} as well as one- \cite{mok01,ang02} and two-nucleon 
\cite{ang03,ang04} emission processes induced by both
electrons and real photons, providing a unified and consistent
description of all these processes.

The linear response of the nucleus 
to an external operator $O(\bqu)$ can be written as
\beq
S(\bqu,\omega) \, = \,
- \frac{1}{\pi} \, \rm{Im} \, \left[
\sum_n \, \xi_n^+(\bqu)
\, (E_n - E_0 - \omega + i \eta)^{-1} \, \xi_n(\bqu)
\right] \, ,
\eeq
where we have defined 
\beq
\xi_n(\bqu) \, = \, \frac { \langle \Psi_n | \,O(\bqu)\, |\Psi_0 \rangle
 } {\langle \Psi_n |\Psi_n\rangle ^{\half}\,\langle \Psi_0
 |\Psi_0\rangle ^{\half} }
\eeq
This function involves the transition matrix element between the
initial and the final states of the nucleus.  These states are
constructed by acting with a correlation operator $F$ on uncorrelated
Slater determinants, as established by the Correlated Basis Function
theory: $|\Psi_0\rangle \,= \,F \,|\Phi_0\rangle$ and 
$|\Psi_n\rangle \,= \,F \,|\Phi_n\rangle$.

In our model we have considered only scalar correlation functions of the
type 
%
%$F = \displaystyle \prod_{i<j}\, f_{ij}$,
\beq 
F \,=\, \displaystyle \prod_{i<j}\, f_{ij} \, , 
\eeq
which commute with the transition
operator $O(\bqu)$. Thus the $\xi$ function can be written as
\beq
\xi_n(\bqu) \, = \, \frac {\langle
\Phi_n |\,O(\bqu)\, \displaystyle 
\prod_{i<j}\,(1+h_{ij}) \,|\Phi_0 \rangle } {\langle
\Phi_0| \displaystyle \prod_{i<j}\,(1+h_{ij}) \,|\Phi_0\rangle }\, 
\left[ \frac {\langle
\Phi_0 | \displaystyle \prod_{i<j}\,(1+h_{ij}) \,|\Phi_0\rangle }
{\langle \Phi_n| \displaystyle \prod_{i<j}\,(1+h_{ij}) \, |\Phi_n\rangle } 
\right]^\half \, , 
\eeq
where $h_{ij}= f^2_{ij}-1$. We assume that the operator $O(\bqu)$
which induces the transitions is a one-body
operator. We perform the full cluster expansion of this
expression and the presence of the denominator
allows us to eliminate the unlinked diagrams. At
this point we insert the main approximation of our model by truncating
the resulting expansion such as only those diagrams which involve a
single correlation function $h$ are retained
\beq
\xi_n(\bqu) \rightarrow \xi^1_n(\bqu) \,=\,
\langle \Phi_n |\,O(\bqu)\, 
( 1+\sum_{i<j}\, h_{ij} ) \,|\Phi_0 \rangle _L \, .
\eeq
In the above expression the subscript $L$ indicates that only linked
diagrams are included in the expansion. 

\begin{figure}
\begin{center}
\epsfig{figure=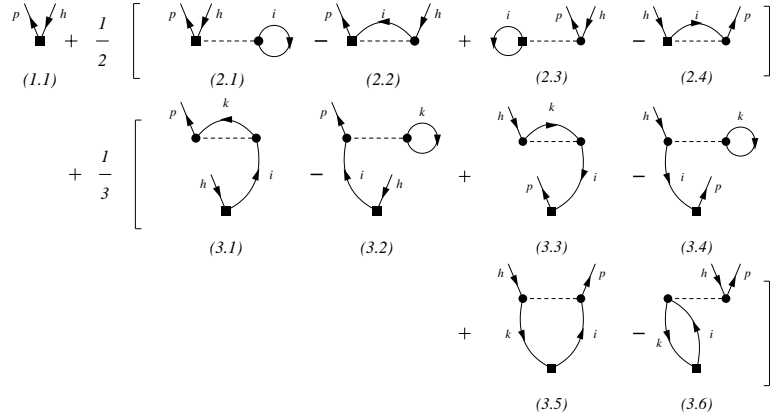,width=10.5cm}
\end{center}
\vspace*{-4mm}
\caption
{\small Meyer-like diagrams contributing to $\xi^1_{\rm 1p1h}(\bqu)$.}
\end{figure} 

\begin{figure}[b]
\begin{center}
\epsfig{figure=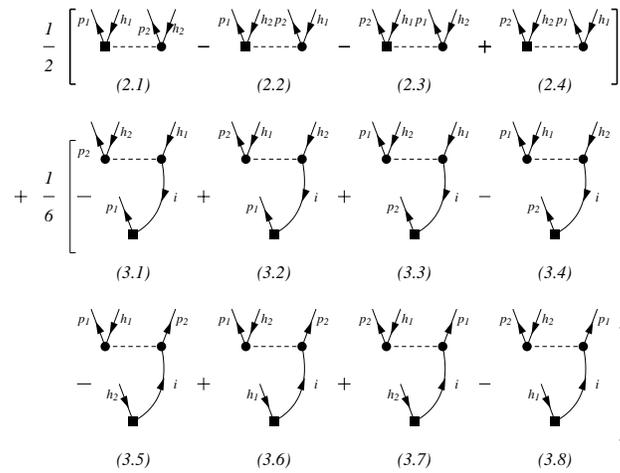,width=8.5cm}
\vspace*{-4mm}
\end{center}
\caption
{\small Meyer-like diagrams contributing to 
$\xi^1_{\rm 2p2h}(\bqu)$. }
\end{figure}

For one-particle one-hole final states, the function
$\xi$ includes three terms:
\beq
\xi^1_{\rm 1p1h}(\bqu) \, = \, 
\langle \Phi_{\rm 1p1h} |\, O(\bqu)\, |\Phi_0 \rangle
+ \, \langle \Phi_{\rm 1p1h} |\, O(\bqu) \, 
\displaystyle
\sum^{A}_{j>1}\, h_{1j}\, |\Phi_0 \rangle
+ \, \langle \Phi_{\rm 1p1h} |\, O(\bqu) \, 
\displaystyle
\sum^{A}_{1< i <j } \, h_{ij} \, |\Phi_0 \rangle \, .
\eeq
The contributions of the various terms of the above expression are
shown in Fig. 1 in terms of Meyer-like diagrams. In addition to the
uncorrelated transition represented by the one-point diagram (1.1),
also four two-point (2p) diagrams and six three-point (3p) diagrams
are present. All these terms are necessary to get a proper
normalization of the nuclear wave functions. In the figure, the black
squares represent the points where the external operator is acting,
the dashed lines represent the correlation function $h$ and the
continuous oriented lines represent the single-particle wave
functions. The letters $h$, $i$ and $k$ label holes, while $p$ labels
a particle. A sum over $i$ and $k$ is understood.

In the case one has two-particle two-hole final states, the function $\xi$
includes two terms:
\beq
\xi^1_{\rm 2p2h}(\bqu) \, = \, \langle \Phi_{\rm 2p2h} |\,O(\bqu) \,
\displaystyle
\sum^{A}_{1<j}\,h_{1j}\, |\Phi_0 \rangle \, + \, 
\langle \Phi_{\rm 2p2h} |O(\,\bqu) \,
\displaystyle
\sum^{A}_{1< i  <j } \,h_{ij}\, |\Phi_0 \rangle \, .
\eeq
The first term consists of four 2p diagrams. The second one results
from the sum of eight 3p diagrams. As in the previous case, this set
of diagrams conserves the correct normalization of the nuclear wave
functions. 

Our model has been tested by comparing our results for the nuclear matter
charge response functions with those obtained by considering the full
cluster expansion. In these calculations the correlation used has been
the scalar part of a complicated state dependent correlation fixed to
minimize the binding energy in a Fermi Hypernetted chain calculation
with the Urbana V14 NN potential \cite{wirin}.  We have found
\cite{ama98} an excellent agreement between the results of the two calculations, 
and this gave us confidence to extend the model to other situations.

\begin{figure}
\begin{center}
\epsfig{figure=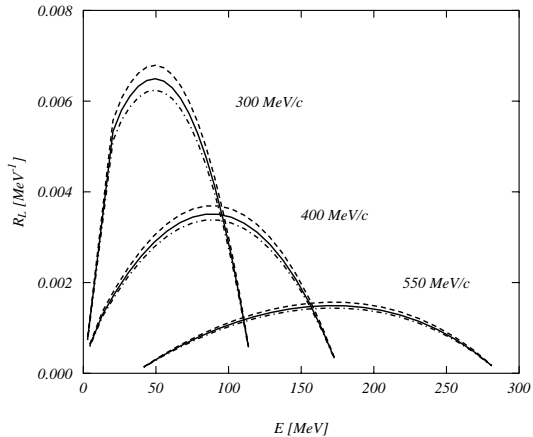,width=7cm}
\end{center}
\vspace*{-7mm}
\caption
{\small Nuclear matter longitudinal responses for $q=300$, 400
and 550~MeV/$c$ and $k_F=1.09$~fm$^{-1}$. Dashed lines represent the
Fermi gas responses, the dashed-dotted lines have been obtained adding
the two-point diagrams, while the full lines show the results of the
complete calculations where all diagrams of Fig. 1 have been
included. The nucleon form factors have been taken from
Ref. \protect\cite{hoel}.}
\end{figure}

\begin{figure}[b]
\begin{center}
\epsfig{figure=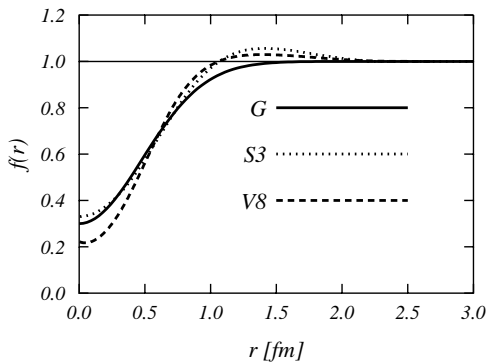,width=7cm}
\vspace*{-6mm}
\end{center}
\caption
{\small Correlation functions used in our calculations.}
\end{figure}

In Fig. 3 we show the effect of the different diagrams shown in Fig. 1. 
The dashed curves represent the uncorrelated Fermi gas result. The
dashed-dotted curves show the effect of adding the 2p diagrams and,
finally, the full curves give the final result after including the 3p
diagrams. One can appreciate how 2p and 3p diagram contributions
interfere destructively. This is a general
characteristic of all our results.

In order to determine the sensitivity of our results to the details of
the SRC we have considered three different correlation functions which
are shown in Fig. 4.  Two corrrelations, labeled $G$ and $S3$, are
fixed by minimizing the energy functional calculated with a nuclear
hamiltonian containing the Afnan and Tang NN potential \cite{ari96}.
The first correlation is of gaussian type, $f(r) = 1 - a \exp \left(
-b r^2 \right)$, and the minimization is done with respect to $a$
and $b$, obtaining $a$=0.7 and $b$=2.2~fm$^{-2}$. The $S3$ correlation
is determined by using the Euler procedure, in which the minimization
is carried out with respect to a single parameter, the healing
distance.  The $V8$ correlation is the scalar part of a state
dependent correlation function fixed again with the Euler procedure
but for a hamiltonian which includes the NN Argonne $V8'$ plus the
Urbana IX three body potentials \cite{fab00}.

Despite the small differences observed between the three correlation
functions, we shall show that, in various cases, they can produce 
rather different results.

\section{Results}

\subsection{Inclusive (e,e') processes}

We start our study with the inclusive (e,e') processes. Here we have
investigated both the excitation of low-lying high-spin states and the
quasi-elastic peak. In these calculations we have considered only the
one-body charge and current operators. 

The excitation of low-lying nuclear excited states was investigated to
test the claim of Pandharipande, Papanicolas and Wambach
\cite{pandha84}. These authors ascribed to SRC the quenching shown by
the experimental data with respect to the calculations. It was argued
that SRC produce the partial occupation of the single particle states
around the Fermi level and this gives rise to the reduction in the
cross sections.

\begin{table}[b]
\begin{center}
\begin{tabular}{ccrcrrrr}
\hline\hline
 & J$^{\pi}$ & $E$ (MeV) & p-h pairs &
\multicolumn{2}{c}{IPM}&\multicolumn{2}{c}{Correlated}\\ 
       & & & &  $Q$ & $\chi^2$ & $Q$ & $\chi^2$ \\
\hline
\rule{0cm}{.3cm}$^{16}$O & 4$^-$ & 18.98 & $\pi$ 1d$_{5/2}$ 1p$_{3/2}^{-1}$ 
&&&& \\[-.2cm]
&&&& 0.67 & 26.35 & 0.66 & 35.06 \\[-.2cm]
         &  &     &$\nu$ 1d$_{5/2}$ 1p$_{3/2}^{-1}$ & & & & \\
\hline
\rule{0cm}{.3cm}$^{208}$Pb
 & 9$^+$ & 5.01   &$\nu$ 2g$_{9/2}$ 1i$_{13/2}^{-1}$ & 0.38 & 5.92 & 0.37 & 7.17  \\
 & 9$^+$ & 5.26   &$\pi$ 1h$_{9/2}$ 1h$_{11/2}^{-1}$ & 0.59 & 2.09 & 0.59 & 2.10  \\
 &10$^-$ & 6.283  &$\nu$ 1j$_{15/2}$ 1i$_{13/2}^{-1}$& 0.34 & 9.12 & 0.34 & 9.29 \\
 &10$^-$ & 6.884  &$\pi$ 1i$_{13/2}$ 1h$_{11/2}^{-1}$& 0.33 & 6.05 & 0.33 & 6.05 \\
 &12$^-$ & 6.437  &$\nu$ 1j$_{15/2}$ 1i$_{13/2}^{-1}$& 0.70 & 3.35 & 0.68 & 3.57 \\
 &12$^-$ & 7.064  &$\pi$ 1i$_{13/2}$ 1h$_{11/2}^{-1}$& 0.28 & 7.64 & 0.27 & 8.85 \\
 &14$^-$ & 6.745  &$\nu$ 1j$_{15/2}$ 1i$_{13/2}^{-1}$& 0.39 & 5.53 & 0.39 & 5.53 \\
{$\theta=90^0$}
 &10$^+$ & 5.920  &$\nu$ 1i$_{11/2}$ 1i$_{13/2}^{-1}$& 0.63 & 18.90 & 0.69 & 20.63 \\
{$\theta=160^0$}
 &10$^+$ & 5.920  &$\nu$ 1i$_{11/2}$ 1i$_{13/2}^{-1}$& 0.88 & 28.91 & 0.95 & 32.15 \\
{$\theta=90^0$} 
 &12$^+$ & 6.100  &$\nu$ 1i$_{11/2}$ 1i$_{13/2}^{-1}$& 0.52 &  7.55 & 0.57 & 8.76 \\
{$\theta=160^0$} 
 &12$^+$ & 6.100  &$\nu$ 1i$_{11/2}$ 1i$_{13/2}^{-1}$& 0.39 & 12.70 & 0.42 & 14.84  \\ 
\hline\hline
\end{tabular}
\end{center}
\vspace*{-5mm}
\caption
{\small Quenching factors obtained for the different excited
states. The excitation energies $E$, the particle-hole pairs
considered and the quenching factors $Q$ and the $\chi^2$ obtained for
the calculations without and with SRC are given. In the case of the
electric states the quenching factors have been obtained for the full
cross section and the corresponding scattering angles have been
indicated.}
\end{table}

The effect of SRC as calculated in our model is very
small. In order to quantify it, we have determined the quenching
factor for different excited states. It is defined as the 
multiplying factor
needed to bring the theoretical results in agreement with the
experimental 
data. The values obtained by minimizing the corresponding
$\chi^2$ are shown in Table 1. As we can see, the values obtained with
and without SRC are very similar. Only in the cases analyzed for
electric states (last four rows), the quenching factors increase by
more than 5\% after SRC are included. However, the values obtained are
still far from 1.  In these calculations the $S3$ correlation function
has been used.

It is interesting to note the cases of the high-spin magnetic states
in $^{208}$Pb, $12^-$ and $14^-$, which were those quoted by
Pandharipande {\it et al.} in \cite{pandha84}. We see here that they are almost
insensitive to the SRC.

Now we analyze the (e,e') process in the quasi-elastic peak. The
calculations we have performed for $^{16}$O and $^{40}$Ca \cite{co01}
have shown that the SRC produce very small effects in both the
longitudinal and the transverse responses. The maximum variation in
the peaks is around 2\%, certainly within the range of the
uncertainties of the calculations (for example, those due to the
nucleon form factor choice).

In Fig. 5 we show the differences between correlated 
and uncorrelated responses after including the 2p (solid curves) and 
also the 3p (dashed curves) diagrams, for both the $S3$ and $V8$
correlations. The left (right) panels of the figure 
show the longitudinal
(transverse) responses.  Three values of the momentum transfer have
been considered. As we can see, the inclusion of the 3p diagrams
always reduces the effects produced by the 2p diagrams alone.

\begin{figure}
\begin{center}
\epsfig{figure=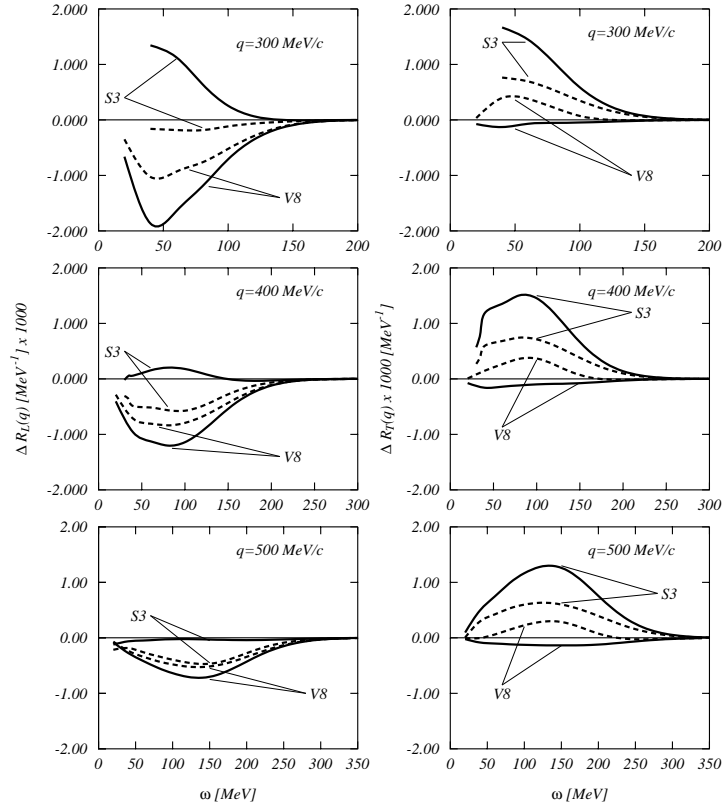,width=9.5cm}
\end{center}
\vspace*{-7mm}
\caption
{\small Differences between correlated and uncorrelated
responses for $^{16}$O(e,e'), after including the 2p (solid curves)
and also the 3p (dashed curves) diagrams, for both the $S3$ and $V8$
correlations, for the longitudinal and the transverse responses and
for 300, 400 and 500~MeV/$c$.}
\end{figure}

In addition, it is interesting to point out the different behavior shown
by the two correlation functions used here. The 2p contributions are
positive for $S3$ and negative for $V8$ and the total contribution in
case of the $S3$ function is always larger than that of the $V8$.

Despite this, it is worth to notice that SRC reduce the longitudinal
response and increase transverse one. This is due to the fact
that the diagram 2.3 of the 1p1h term (see Fig. 1) 
does not contribute to the transverse response.

\subsection{One-nucleon emission}

Now we discuss the main results we have obtained for one-nucleon
emission processes \cite{mok01,ang02}.  In these calculations, in
addition to the usual one-body charge and current operators, one has
to consider also the Meson Exchange Currents (MEC). Our analysis of
the role of these currents \cite{ang02} shows that their contributions
become small for momentum transfer values larger than 300 MeV/c. For
this reason we neglect them in (e,e'p) processes. On the contrary, MEC
are relevant in photo-emission processes.  Specifically, in this case,
we have considered the so-called seagull, pion-in-flight and
$\Delta$-isobar currents.

In the nucleon emission knock-out processes,
the wave functions of the hole states have
been described by using a real Woods-Saxon potential, and those of
the emitted particles by an optical potential \cite{Schw}.

\begin{figure}
\begin{center}
\epsfig{figure=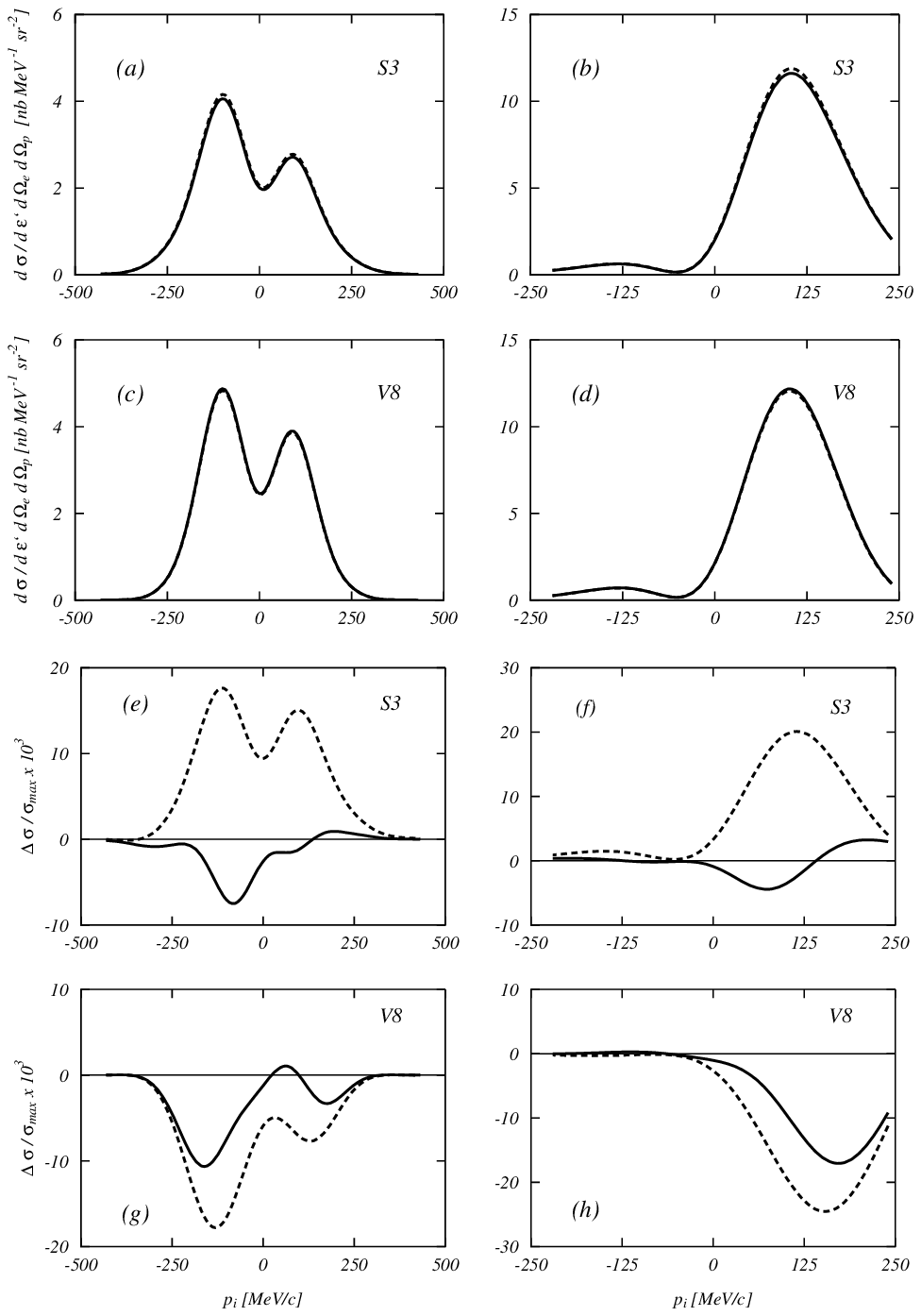,width=9.5cm}
\end{center}
\vspace*{-8mm}
\caption
{\small Panels (a)-(d) show the cross sections of the $^{16}$O(e,e'p)
process as a function of the nucleon initial momentum $p_i=|{\bf
p}-{\bf q}|$. The proton is emitted from the $1p_{1/2}$ level. The
cross sections have been calculated for $\omega=128$~MeV and $|{\rm
p}|=444$~MeV/$c$. Left (right) panels correspond to perpendicular
(parallel) kinematics. The dashed curves include the uncorrelated and the
2p diagrams. The full curves include also the 3p diagrams. 
The panels (e)-(h)
show the differences between the correlated and uncorrelated cross
sections, divided by the maximum values of the uncorrelated cross
sections, $\sigma_{\rm max}$. Results for both the $S3$ and $V8$
correlations are shown.}
\end{figure}

Panels (a)-(d) of Fig. 6 show the cross section for the
$^{16}$O(e,e'p) process when the proton is emitted from the 1p$_{1/2}$
hole, for perpendicular (left panels) and parallel (right panels)
kinematics. Calculations have been done for fixed values of
$\omega=128$~MeV and $|{\rm p}|=444$~MeV/$c$.  The dashed curves show
the results obtained by including the 2p diagrams, while the full
curves are obtained if 3p diagrams are added. In the scale of the
figures, these last curves overlap almost exactly with those obtained
without SRC (which are not shown in the figure). The effects of the
SRC are always very small.  In order to emphasize these effects, we
show in the panels (e)-(h) of Fig. 6 the differences between
correlated and uncorrelated cross sections, of the above calculations.
We see again that the cross sections obtained for the $S3$ and $V8$
correlations show a very different behavior if only the 2p
contributions are included, while the addition of the 3p diagrams
produces rather similar results in both cases.  This basic results
does not change too much in other kinematic conditions, for example if
the proton is emitted from the 1p$_{3/2}$ or 1s$_{1/2}$ hole state, or
if different values of the momentum
transfer are considered \cite{ang02}.

\begin{figure}
\begin{center}
\epsfig{figure=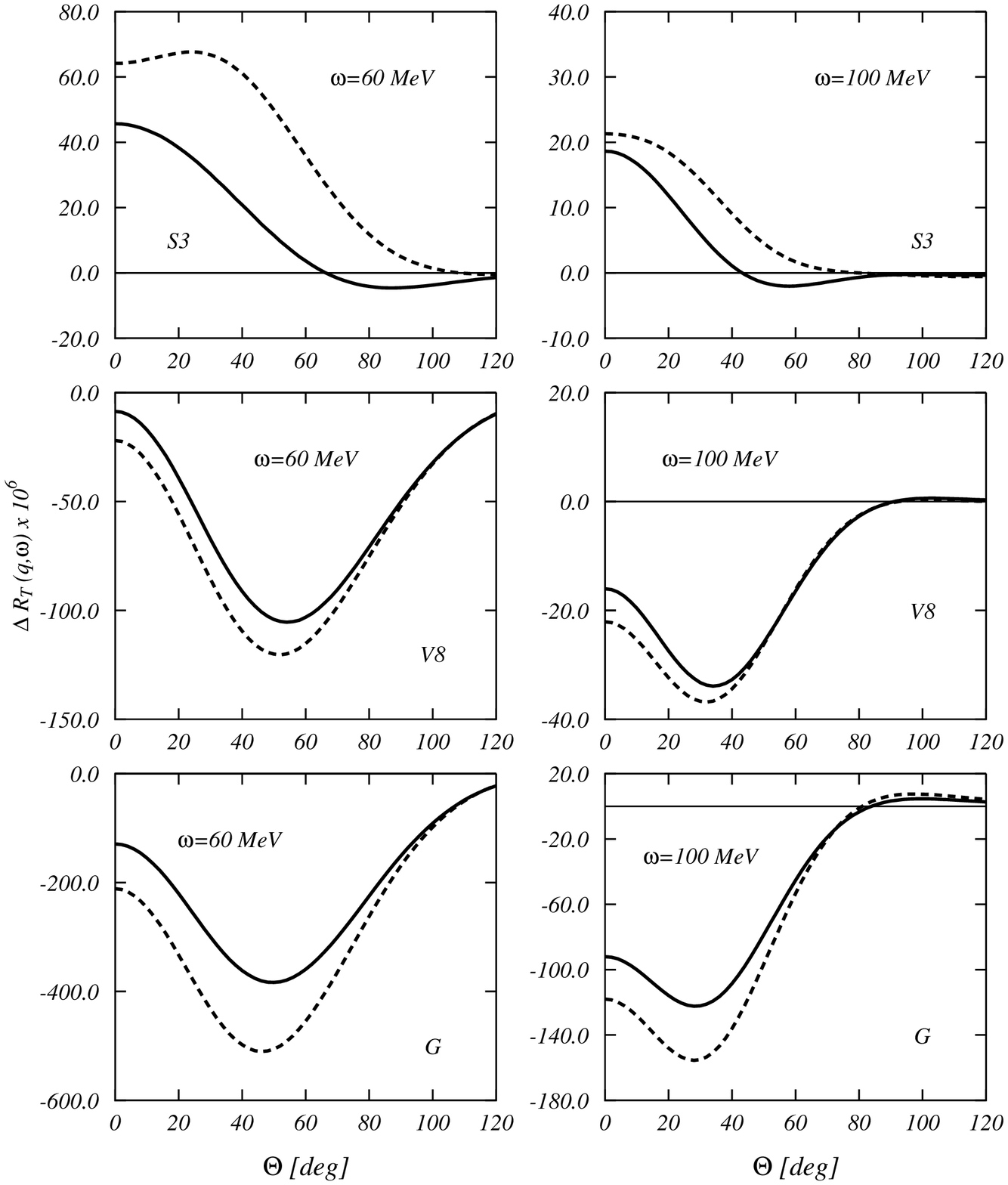,width=9cm}
\end{center}
\vspace*{-8mm}
\caption
{\small Relative differences between correlated and
  uncorrelated cross sections, for the $^{16}$O($\gamma$,p) process,
  for the three correlation functions considered and for two values of
  the energy
  transfer $\omega$. The proton is emitted from the $1p_{1/2}$ level.}
\end{figure}

\begin{figure}
\begin{center}
\epsfig{figure=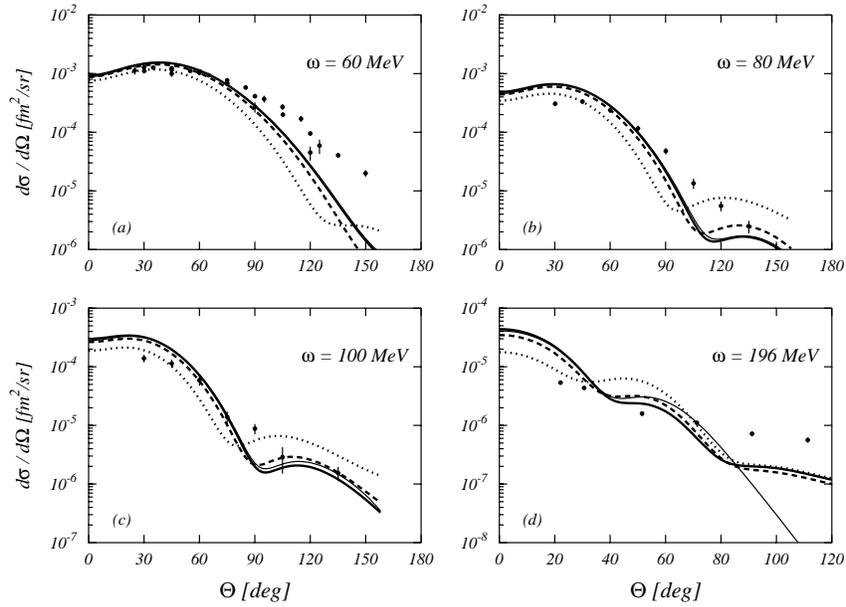,width=8.cm,angle=90}
\end{center}
\vspace*{-6mm}
\caption
{\small Angular distributions for the $^{16}$O($\gamma$,p)
process, calculated without SRC (thin full curves) and with the $S3$
(thick full curves), $V8$ (dashed curves) and $G$ (dotted curves)
correlations. The proton is emitted from the $1p_{1/2}$ level.}
\end{figure}

\begin{figure}
\begin{center}
\epsfig{figure=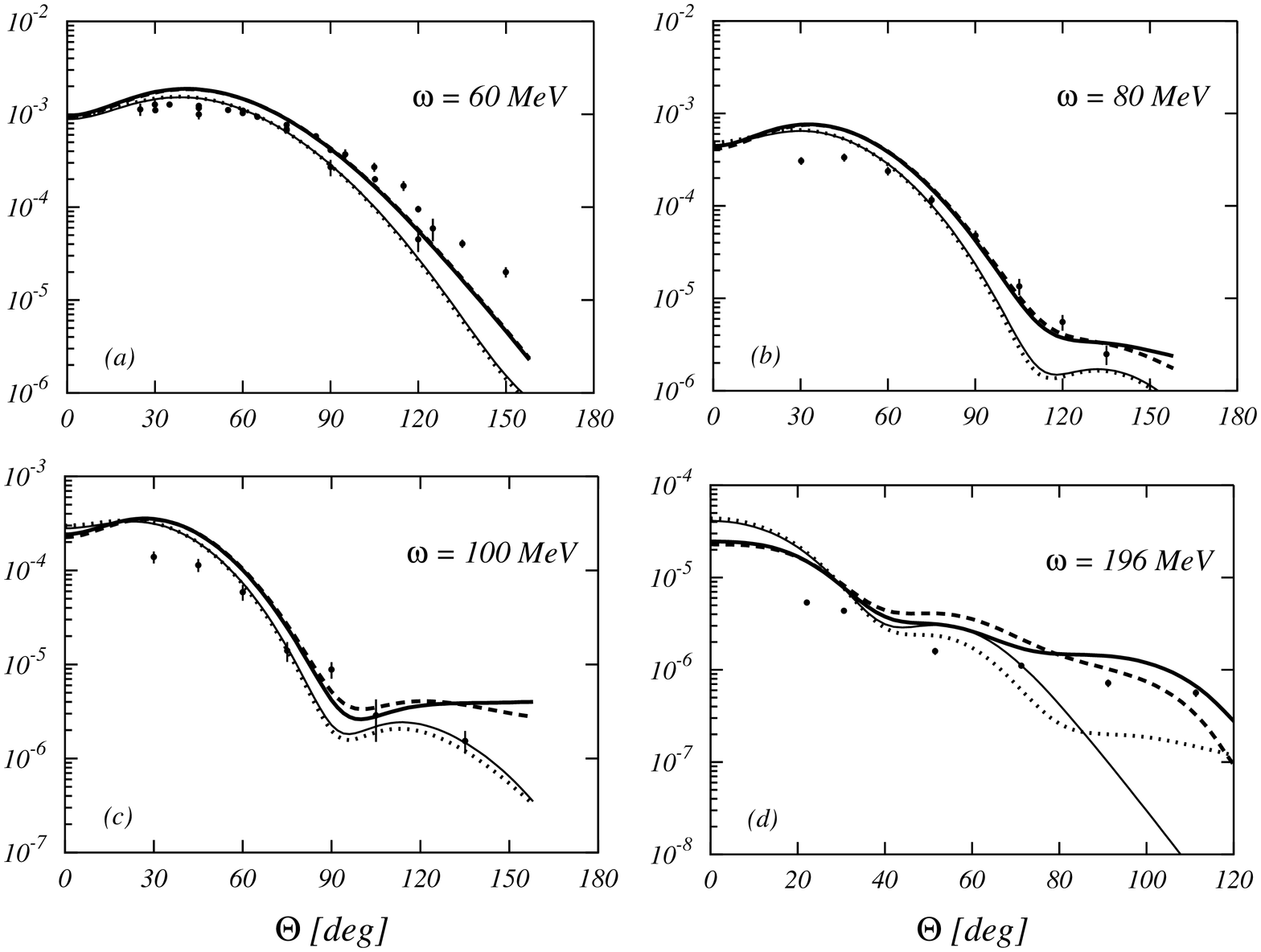,width=8.cm,angle=90}
\end{center}
\vspace*{-6mm}
\caption
{\small Angular distributions of the $^{16}$O($\gamma$,p)
process, calculated without SRC and MEC (thin full curves), by adding
only the $S3$ correlation (dotted curves), by adding only the MEC
(dashed curves) and with all the contributions (thick full curves),
one-body plus SRC plus MEC. The proton is emitted from the $1p_{1/2}$
level.}
\end{figure}

Now we analyze the photoemission process. Fig. 7 shows, for the
$^{16}$O($\gamma$,p) process, the relative differences between
correlated and uncorrelated responses when only the 2p diagrams are
included (dashed curves), and when also the 3p diagrams are considered
(solid curves). Again, one can see how the contributions of the 3p
diagrams cancel those due to the 2p ones in all cases.  However, the
full effect of the SRC for the $S3$ correlation (upper panels) shows a
different sign with respect to 
that obtained with the $V8$ (middle panels) and 
with the $G$ (lower panels) correlations. Furthermore we point out
that the order of magnitude of the correlation effects
obtained with the $G$ correlation 
are about a factor $4$  larger than those obtained with the other
correlation functions.

Fig. 8 shows the angular distribution for the $^{16}$O($\gamma$,p)
calculated for various incident photon energies.  In this figure, the
uncorrelated cross sections, which are shown by the thin full curves,
are compared with those obtained by using the $S3$ (thick full
curves), $V8$ (dashed curves) and $G$ (dotted curves) correlation
functions. A comparison between the SRC effects presented in this figure
and those presented in Fig. 6 shows that the ($\gamma$,p) process has a larger
sensitivity to the SRC functions that the (e,e'N).

As we have already mentioned, MEC effects are not
negligible in photoemission reactions. 
In Fig. 9 we show the same cross sections of Fig. 8, but
now with the MEC included.
In Fig.9 the thin full lines represent the results obtained
using the one-body currents only, i.e. 
without considering SRC and MEC. The addition of SRC (dotted curves)
produces a modification much smaller than that found if only MEC are
included (dashed curves). Thick solid curves correspond to the full
results and it is evident that MEC dominate even in the large angle
region, where SRC effects show up in a clearer way. 

\subsection{Two-proton emission}

We finally discuss the two-proton emission processes induced by
electrons \cite{ang03} and photons \cite{ang04}. In these processes
the bare one-body currents contribution is absent. The one-body
currents act only if linked to the SRC. When two-like particles are
emitted, in our case two protons, the MEC produced by the exchange of
charged mesons are not active. This means that in the two-proton
emission, the only MEC competing with the SRC are $\Delta$ current due
to the exchange of a chargeless pion.  The description of the single
particle wave functions is the same as that adopted for the single
nucleon knock out.  We have studied two-proton knock out from
$^{12}$C, $^{16}$O and $^{40}$Ca. The final states of the residual
nuclei considered in our calculations are given in Table 2.

\begin{table}[b]
\begin{center}
\begin{tabular}{cccc}
\hline\hline
         &   $^{12}$C     &  $^{16}$O      & $^{40}$Ca \\
\hline
 $0_1^+$ & (1p3/2)$^{-2}$ & (1p1/2)$^{-2}$ & (1d3/2)$^{-2}$  \\
 $0_2^+$ &                & (1p3/2)$^{-2}$ & (2s1/2)$^{-2}$  \\
 $1^+$   &                & (1p1/2)$^{-1}$  (1p3/2)$^{-1}$ &  
                                (1d3/2)$^{-1}$ (2s1/2)$^{-1}$ \\
 $2_1^+$ & (1p3/2)$^{-2}$ & (1p1/2)$^{-1}$  (1p3/2)$^{-1}$ &
                                           (1d3/2)$^{-2}$ \\
 $2_2^+$ &                & (1p3/2)$^{-2}$ &  (1d3/2)$^{-1}$
                                              (2s1/2)$^{-1}$  \\   
\hline\hline
\end{tabular}
\end{center}
\vspace*{-5mm}
\caption
{\small Two-hole compositions of the nuclear final states of 
the residual nuclei for the two proton emission process we have considered. }
\end{table}

\begin{figure}
\begin{center}
\epsfig{figure=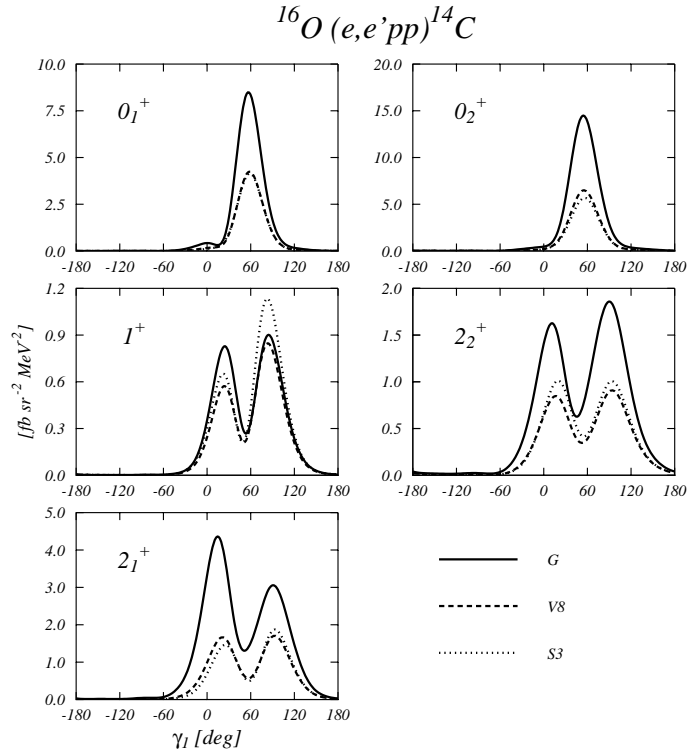,width=9cm}
\end{center}
\vspace*{-6mm}
\caption
{\small Angular distributions of the (e,e'pp) cross
sections for $^{16}$O, for different final states (see Table 2) and
for the three correlation functions.}
\end{figure}

\begin{figure}
\begin{center}
\epsfig{figure=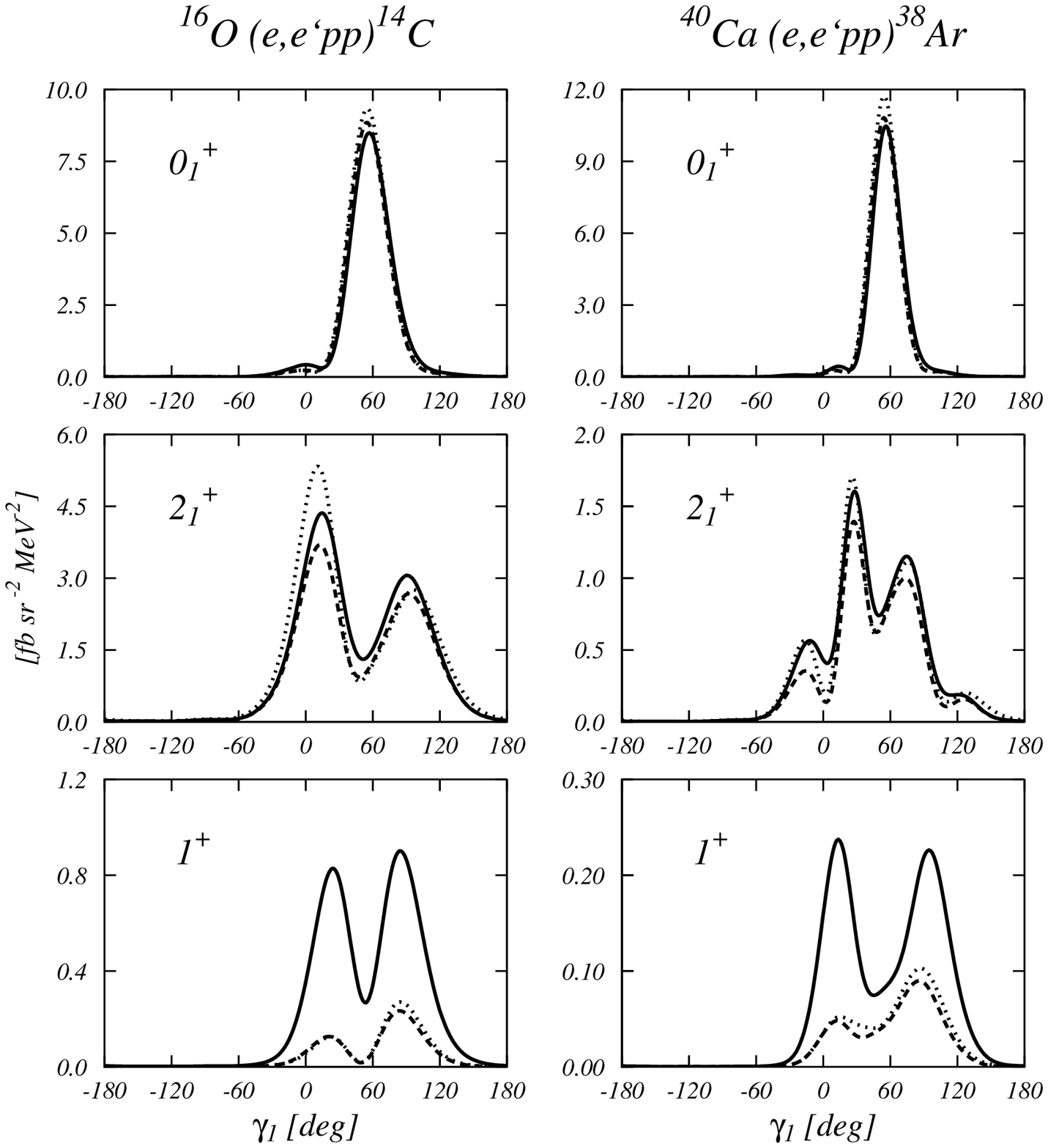,width=10.cm}
\end{center}
\vspace*{-6mm}
\caption
{\small (e,e'pp) cross sections for $^{16}$O (left panels)
and $^{40}$Ca (right panels). The dotted curves have been obtained
with 2p diagrams only. The dashed lines show the results obtained by
adding 3p diagrams, and the solid lines those obtained by including
also the $\Delta$ currents. The gaussian correlation has been used.}
\end{figure}

In Fig. 10 we show the (e,e'pp) cross sections for $^{16}$O,
calculated for different final states with the three correlation
functions shown in Fig. 4. In Fig. 10 the cross sections are shown as
a function of the emission angle of one of the emitted protons. We
have assumed coplanar kinematics, and we fixed the energy and emission
angle of the other proton at 40~MeV and $60^o$ respectively.
The incident electron energy has been chosen to be 800~MeV and the
energy and momentum transferred to the nucleus have been fixed at 100~MeV and
400~MeV/$c$, respectively.
The results of Fig. 10 show that the SRC modify the size of the cross
sections, while the shape is maintained. The cross section obtained
for the $S3$ and $V8$ are a factor 2 smaller than those found for the
$G$ correlation. The $1^+$ is out of the systematic because it is
dominated by the $\Delta$ current. Similar results are
obtained for $^{12}$C and $^{40}$Ca \cite{ang03}.\\

The fact that the $\Delta$ current dominates the $1^+$ cross sections
is evident also from the results shown in Fig. 11, where the cross
sections obtained with the gaussian correlation for $^{16}$O and
$^{40}$Ca are presented for three different final states. These
calculations have been done for the same kinematics of Fig. 10.
In Fig. 11, the dotted lines show the results obtained with the 2p
diagrams only, while the dashed and solid lines have been obtained by
adding first the 3p diagrams and then the $\Delta$ currents.  The 3p
diagrams reduce the cross sections obtained with the 2p diagrams only.
This reduction depends from the angle of the emitted proton.  This is
particularly clear in the case of the $2^+$ states.

The last process we discuss is ($\gamma$,pp). In Fig. 12 we show the
angular distributions of the cross sections corresponding to the three
nuclei we are investigating. We have considered the case when
the remaining nucleus is left in its ground state. 
The results have been obtained by using the three correlation
functions of Fig. 4. Again, the main differences are in the size of
the cross sections.

\begin{table}
\begin{center}
\begin{tabular}{ccccccc}
\hline\hline
 &\multicolumn{2}{c}{$^{12}$C} &  \multicolumn{2}{c}{$^{16}$O} & 
         \multicolumn{2}{c}{$^{40}$Ca} \\
\hline
         & $S3$  & $V8$     & $S3$ & $V8$     & $S3$ & $V8$ \\
\hline
 $0_1^+$  & 0.73 $\pm$ 0.11 & 0.52 $\pm$ 0.05  
          & 0.10 $\pm$ 0.11 & 0.20 $\pm$ 0.05 
          & 0.15 $\pm$ 0.03 & 0.22 $\pm$ 0.04  \\
 $0_2^+$ & &   
         & 0.83 $\pm$ 0.16 & 0.60 $\pm$ 0.08 
         & 0.37 $\pm$ 0.07 & 0.44 $\pm$ 0.07  \\
 $1^+$   & &   
         & 0.97 $\pm$ 0.08 & 0.76 $\pm$ 0.07
         & 0.37 $\pm$ 0.04 & 0.36 $\pm$ 0.06  \\
 $2_1^+$ & 0.56 $\pm$ 0.13 & 0.52 $\pm$ 0.08  
         & 0.25 $\pm$ 0.05 & 0.30 $\pm$ 0.04 
         & 0.22 $\pm$ 0.05 & 0.24 $\pm$ 0.03  \\
 $2_2^+$ & &   
         & 0.54 $\pm$ 0.13 & 0.52  $\pm$ 0.08 
         & 0.54 $\pm$ 0.07 & 0.48 $\pm$ 0.05  \\
\hline\hline
\end{tabular}
\end{center}
\vspace*{-5mm}
\caption
{\small Ratio of the $S$ factor (see
Eq. \protect\ref{eq:S-fac}) calculated for the $S3$ and $V8$
correlations and that for the $G$ correlation, averaged on the
$\theta_2$ variable.}
\end{table}

In order to have a concise information, the integrated quantities
\beq
S^{\rm (SRC)}\, = \, \displaystyle \int {\rm d}\theta_1 \sin \theta_1
 \frac{{\rm d}^5 \sigma^{\rm (SRC)}(\theta_1)}
{ {\rm d}\Omega_1
{\rm d}\epsilon_2 {\rm d}\Omega_2 }
\label{eq:S-fac}
\eeq
have been calculated for the various final states and for various
emission angles of the second proton. The mean values of the ratios
between the $S$ factors calculated with the $S3$ and $V8$ correlations
and those calculated with the $G$ correlations are given in Table 3.
All the values in the table are smaller than 1 and this means that $G$
correlation produce the largest cross sections.

\begin{figure}
\begin{center}
\epsfig{figure=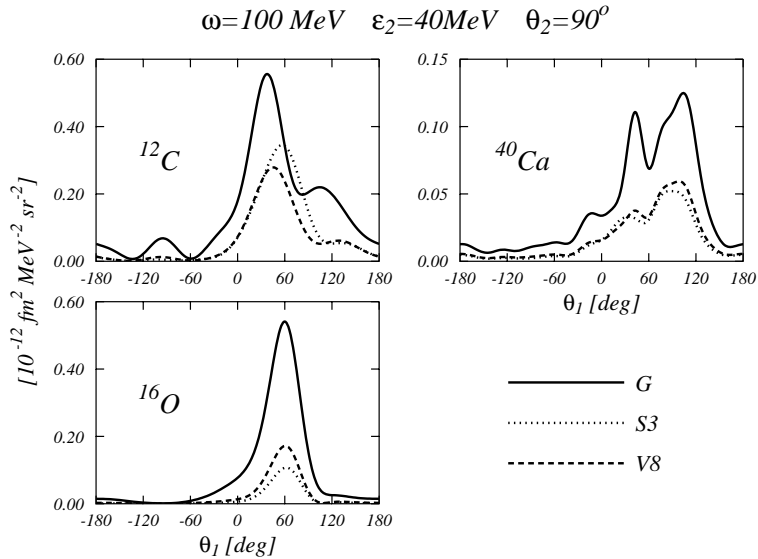,width=11.cm}
\end{center}
\vspace*{-8mm}
\caption
{\small Angular distributions of the ($\gamma$,pp) cross
  sections for the three different nuclei considered. The kinematics
  variables have been fixed as $\omega=100$ MeV, $\theta_2=90^o$ and
  $\epsilon_2$=40 MeV. The final states are the ground states of the
  $A-2$ nuclei, this corresponds to the $0^+_1$ states of Table 2.  The
  full lines have been obtained with the gaussian correlation, the
  dotted ones with the S3 correlation and the dashed lines with the V8
  correlation.}
\end{figure}

\begin{figure}
\begin{center}
\epsfig{figure=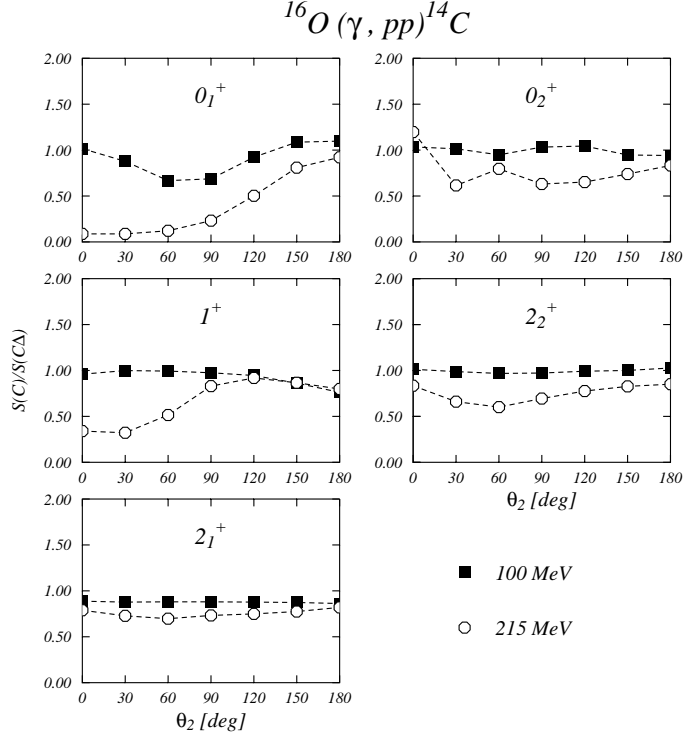,width=10.cm}
\end{center}
\vspace*{-7mm}
\caption
{\small Ratio $S(C)/S(C\Delta)$ (see
Eq. (\protect\ref{eq:S-fac})), calculated for various values of
$\theta_2$ and for photon energies of 100 and 215 MeV. The
calculations have been done with $\epsilon_2$=40 MeV.  The dashed
lines have been drawn to guide the eyes.} 
\end{figure}

\begin{figure}
\begin{center}
\epsfig{figure=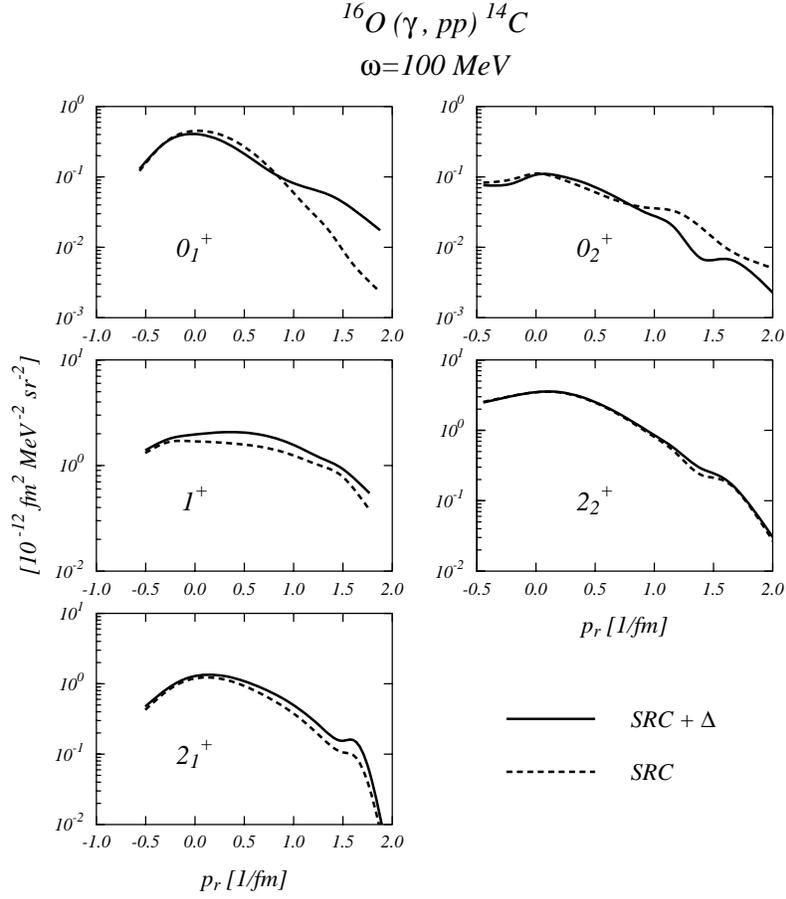,width=11.5cm}
\end{center}
\vspace*{-7mm}
\caption
{\small Cross sections for $^{16}$O ($\gamma$,pp) $^{14}$C
  in superparallel back to back kinematics with $\theta_2=180^o$. The
  full lines have been obtained by considering both SRC and $\Delta$
  terms, while the dashed lines show the results obtained without the
  $\Delta$ terms.}
\end{figure}

\begin{figure}
\begin{center}
\epsfig{figure=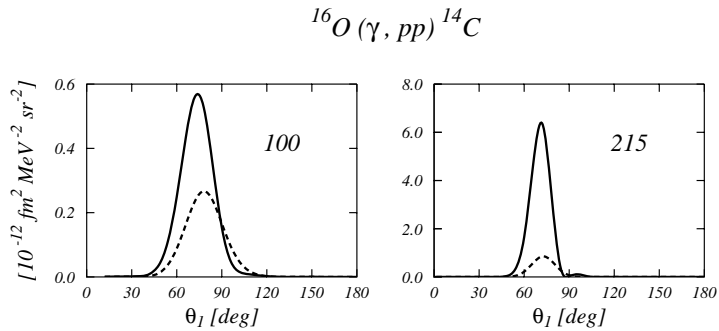,width=10.5cm}
\end{center}
\vspace*{-7mm}
\caption
{\small Angular distributions for $^{16}$O ($\gamma$,pp) $^{14}$C,
  for the $0^+_1$ final state, in symmetric kinematics. The full lines
  show the results obtained by considering both SRC and
  $\Delta$-currents. The results shown by the dashed lines have been
  obtained with SRC only. The numbers characterizing the two panels
  indicate, in MeV, the photon energy.}
\end{figure}

\begin{figure}
\begin{center}
\epsfig{figure=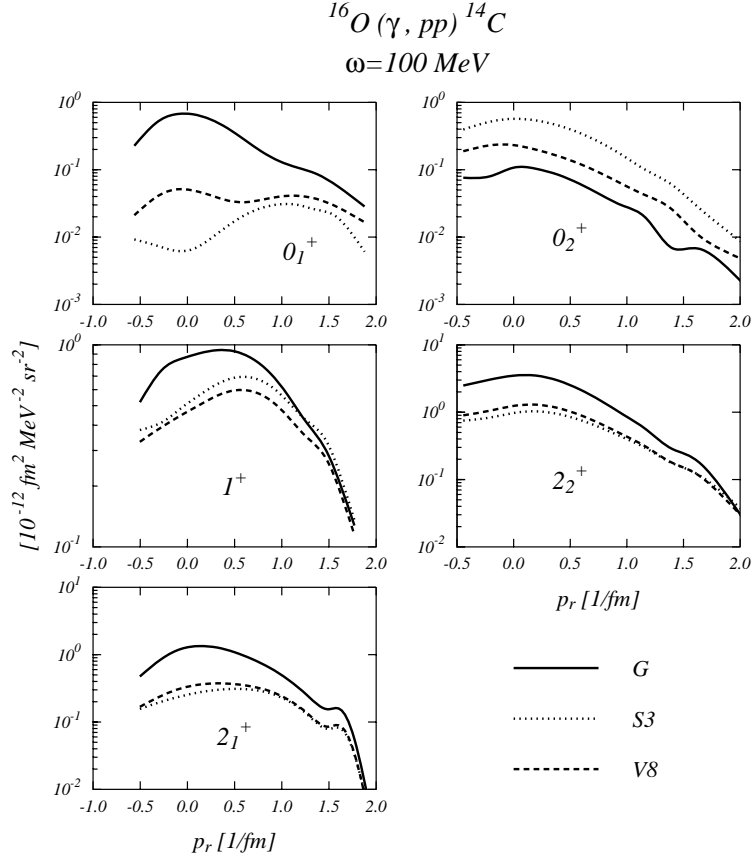,width=11.cm}
\end{center}
\vspace*{-7mm}
\caption
{\small Same as in Fig. 14 for the three correlation functions used.}
\end{figure}

\begin{figure}
\begin{center}
\epsfig{figure=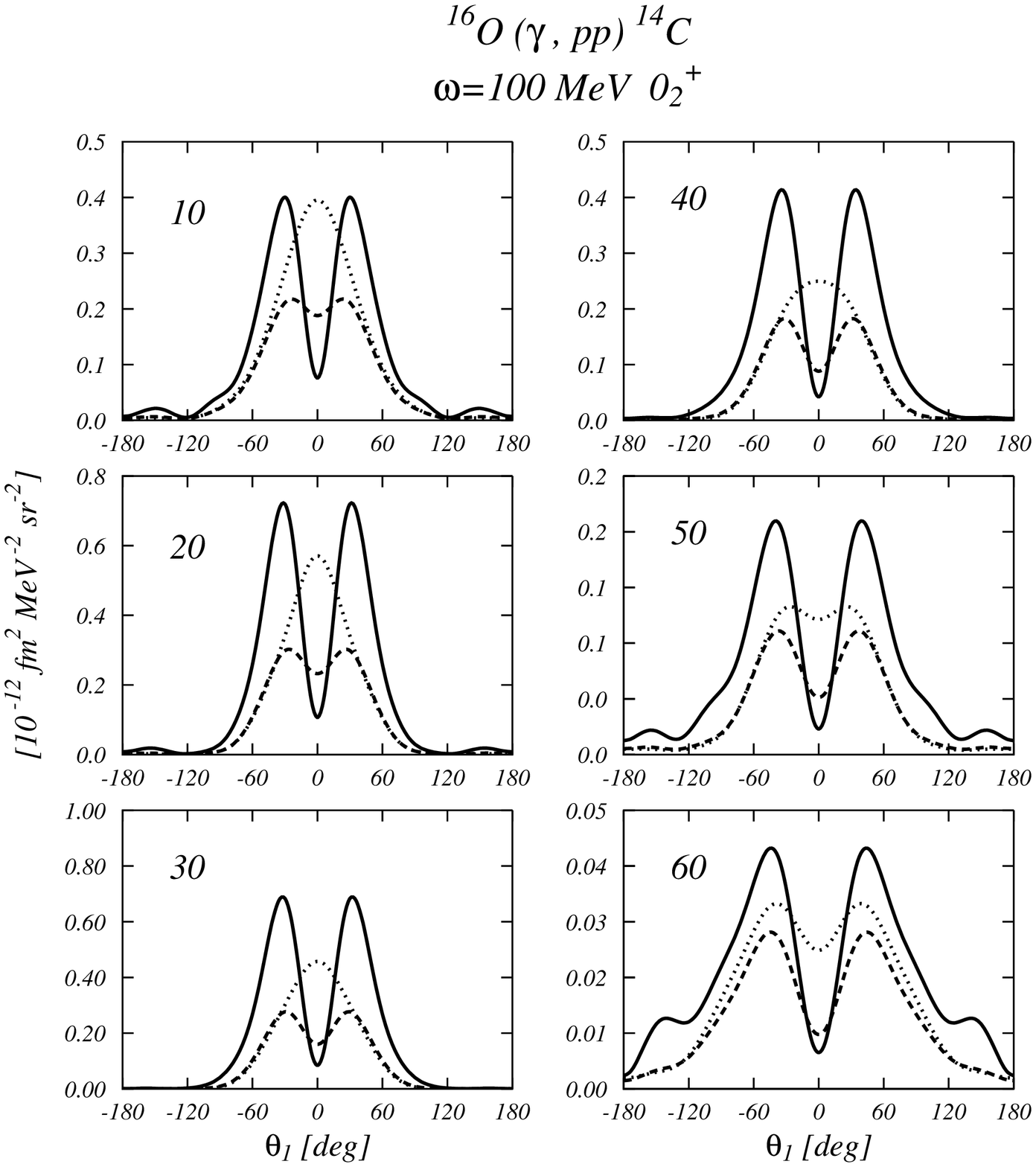,width=10.cm}
\end{center}
\vspace*{-7mm}
\caption
{\small Angular distributions of the $^{16}$O ($\gamma$,pp)
  $^{14}$C cross sections for the $0^+_2$ final state. The numbers in
  the various panels indicate, in MeV, the values of $\epsilon_2$. The
  meaning of the lines is analogous to that of Fig. 16.}
\end{figure}

The second aspect of interest concerns the role of the $\Delta$
current. Fig. 13 shows the ratio between the $S$ factor calculated
with SRC only with that obtained by adding
the $\Delta$ currents contribution as a function of the angle
$\theta_2$ of the second emitted proton. The reaction considered is 
$^{16}$O($\gamma$,pp) and the various panels show the results
obtained for the different final states of the $^{14}$C. 
Two different photon energies have been used.

The effects of the $\Delta$ are smaller at 100 MeV
than at 215 MeV, but they show a strong dependence on the final state
of the residual nucleus and on the proton emission angle.
For $\omega=100$ MeV, all the ratios, except for some cases, are
very close to unity, showing a small effect of the $\Delta$ currents,
even in the case of the $1^+$ state. 
In general, the (e,e'pp) reaction showed larger sensitivity to the
$\Delta$ currents. 

We found analogous results in other kinematics. Fig. 14 shows the
effect of the $\Delta$ in the superparallel back-to-back kinematics.
Again the $\Delta$ appears to have scarce relevance at $\omega=100$
MeV. Nevertheless there is noticeable effect for the $0^+_1$ state at
high values of the recoil momentum.

However, it is possible to find kinematics where the $\Delta$ is
important even at 100~MeV. This is what Fig. 15 shows. In this figure
the results obtained for the $0^+_1$ state in the
$^{16}$O($\gamma$,pp)$^{14}$C reaction in symmetric kinematics are
presented.

In all the results presented so far, those obtained with the gaussian
correlation show the largest SRC effects, as Figs. 10 and 12 indicate. 
This trend regards the full
angular distribution of one of the emitted protons after fixing the
rest of the kinematics. By selecting specific kinematics, in other
words, by choosing some peculiar angular distribution set ups, it is
possible to find situations where this general feature is not present.
An example of this is given in Fig. 16, where the cross sections
calculated for the three different correlation functions in
superparallel back-to-back kinematics are shown. It is evident how the
case of the $0^+_2$ is out of the general trend observed for the
other states.

To understand the results of Fig. 16, we show in Fig. 17 the full
angular distributions of the emitted proton for various energies of
the other proton. The cross sections of the  $0^+_2$ panel in Fig. 16,
have been obtained by selecting the values at $\theta_1=0^o$ in
Fig. 17, and other more. The cross sections obtained with the gaussian
correlation gives the largest integral in all the panels of Fig. 17,
as expected. By selecting the $\theta_1=0^o$ values, one finds the
anomaly of a deep minimum of the full lines, and this produces the
anomalous behaviour shown in  Fig. 16.

\section{Conclusions}

The effects of the SRC in inclusive and one-nucleon emission processes
are rather small, within the uncertainties of the calculations due to
the arbitrary choice of the input parameters. 

In these processes, the uncorrelated one-body responses dominate.
There are situations in ($\gamma$,N) reactions showing certain
sensitivity to the SRC. Unfortunately these are also the situations
where the contribution of MEC is large, even larger than that of the
SRC.

The emission of two nucleons has various advantages, the main one  
is that the one-body uncorrelated terms are absent. However there is
competition between MEC and SRC. The MEC effects can be
reduced by selecting the emission of two-like nucleons. 
In this case the only contribution of the MEC is due to  
the $\Delta$ currents with the exchange of a
chargeless pion.

A priori the longitudinal response of the (e,e'pp) process can provide
a very clean signature of the SRC since the $\Delta$ currents do not
contribute. However, the extraction of this response requires a super
Rosenbluth separation, which is, from the experimental point of view,
a rather difficult procedure.  The alternative is to look for
situations where the $\Delta$ currents produce small effects.

The ($\gamma$,pp) process has a number of advantages with respect to
(e,e'pp).  First, only the transverse response is present.  Second,
the $\Delta$ current effects are minimal at the photon point and, as
we have shown, these effects can be made almost negligible by an
adequate choice of the kinematics.

\section*{Acknowledgments}

We kindly acknowledge the collaboration with Sherif R. Mokhtar, J.
Enrique Amaro, Fernando Arias de Saavedra and Adelchi Fabrocini.

\end{document}